# Multi-level recording in dual-layer FePt-C granular film for heat-assisted magnetic recording

P. Tozman[1*], S. Isogami[1], I. Suzuki[1], A. Bolyachkin[1], H. Sepehri-Amin[1], S.J. Greaves[2] H. Suto[1], Y. Sasaki[1], H.T. Y. Chang[3], Y Kubota[3], P. Steiner[3], P.-W. Huang[3], K. Hono[1] and Y.K. Takahashi[1*]

[1] National Institute for Materials Science, Tsukuba, Ibaraki 305-0047, Japan
[2] Research Institute of Electrical Communications Tohoku University, Sendai, Japan
[3] Seagate Technology, Recording Media Organization, Fremont, CA. 94538 USA

**Abstract**

Multi-level magnetic recording is a new concept for increasing the data storage capacity of hard disk drives. However, its implementation has been limited by a lack of suitable media capable of storing information at multiple levels. Herein, we overcome this problem by developing dual FePt-C nanogranular films separated by a Ru-C breaking layer with a cubic crystal structure. The FePt grains in the bottom and top layers of the developed media exhibited different effective magnetocrystalline anisotropies and Curie temperatures. The former is realized by different degrees of ordering in the $L1_0$-FePt grains, whereas the latter was attributed to the diffusion of Ru, thereby enabling separate magnetic recordings at each layer under different magnetic fields and temperatures. Furthermore, the magnetic measurements and heat-assisted magnetic recording simulations showed that these media enabled 3-level recording and could potentially be extended to 4-level recording, as the ↑↓ and ↓↑ states exhibited non-zero magnetization.

Corresponding authors:
pelin.tozman@tu-darmstadt.de, TAKAHASHI.Yukiko@nims.go.jp

**Evolution of magnetic recording**

The digital transformation from Industry 4.0 to 5.0 has resulted in a significant growth in big data, increasing the demand for data storage [1-3]. However, with the expected increase in energy consumption owing to the construction of more data centers, an environmentally friendly solution is required to satisfy this demand. One solution involves increasing the storage capacity of hard disk drives (HDDs), which currently serve as the primary storage devices in data centers. HDDs comprise a spinning disk coated with a nanogranular ferromagnetic layer (media) and read/write head that magnetically reads and writes data to the disk. The storage capacity of HDDs is determined by the bit size, which refers to the physical dimensions of the region on the media where several ferromagnetic nanograins are magnetized in the same direction. Reducing the bit size facilitates increased data storage in the same physical space [4], thereby increasing the areal density of data bits from the current value of 1.5 Tbit/in$^2$ to 4 Tbit/in$^2$ and beyond [3-8]. This is achieved by reducing the grain size of the granular media while maintaining the long-term thermal stability of the data by satisfying the condition $K_\mathrm{u}V/k_\mathrm{B}T > 60$, where $K_\mathrm{u}$ is the uniaxial magnetocrystalline anisotropy constant, $V$ is the volume of a grain, $k_\mathrm{B}$ is the Boltzmann constant, and $T$ is the temperature [9,10]. This condition is fulfilled by L1$_0$-FePt granular media, where FePt grains with a very high $K_\mathrm{u} \approx 7$ MJ/m$^3$ are uniformly dispersed in a nonmagnetic segregant [7,9,11,12]. However, this medium cannot be used in conventional recording technologies because it requires a high magnetic field, more than 3 T, for the writing process. This problem is overcome via heat-assisted magnetic recording (HAMR), which facilitates the writing of data on high-$K_\mathrm{u}$ and high-coercivity ($H_\mathrm{c}$) media materials [11,13]. Unlike conventional recording, HAMR equips a write head with an optical transducer to heat a small area on the recording media around its Curie temperature $T_c$, thereby lowering the high-$K_u$ locally and enabling the switching of the magnetization direction by a write field generated from the write head (Fig. 1(a)).

According to the Advanced Storage Research Consortium (ASRC), the primary focus for increasing the areal density in HAMR is to improve the nanostructure of the granular FePt-X media [7,9], where X indicates nonmagnetic segregants. By refining the grain size to 4.3 nm and obtaining a high degree of $L1_0$-order, the areal density is expected to reach 4 Tbit/in$^2$ [7,9,10,14,15]. To achieve such a fine microstructure, various X, including metal oxides, B, h-BN, B$_4$C, and C, have been added to FePt-based granular films, with C and h-BN providing a small average grain size, <D>= 5-8 nm while maintaining uniaxial anisotropy [11, 14-20]. Thus, a maximum areal density of 2.77 Tbit/in$^2$ was achieved by Seagate for a demo HAMR [3].

However, FePt grains smaller than 4 nm (<D> < 4 nm) cannot maintain their crystallographic $L1_0$ order, resulting in thermal fluctuations in magnetization owing to their low anisotropy energy [18]. Therefore, increasing the areal density of future generations of HDD recording media requires a new recording concept that is not reliant on grain size miniaturization [8,21]. One solution is bit-pattern media (1 bit = 1 grain) instead of granular media (1 bit = multiple grains). Numerical calculations predicted that bit-patterned media can achieve an areal density of 10 Tbit/in$^2$ [21-23]. However, bit-patterned media cannot be mass produced because of their costly nanofabrication process. Another solution involves the use of multi-level recording [24-26], as explained in the following Section.

**Concept of multi-level recording in HAMR**

A multi-level recording concept has been proposed for HAMR technologies,

which uses media comprising multiple recording layers with different $T_c$ and controls the magnetization reversal in each layer by tuning the temperature of the heated spot [27-29]. Fig. 1 shows a schematic of the dual-layer version of a multi-level HAMR in comparison with the traditional single-layer version of a 2-level HAMR. The single-layer version employs a constant laser power to heat a small region of the magnetic layer around $T_c$, and the write field is used to switch the magnetization direction between the up ↑ ("1") and down ↓ ("-1") magnetization states. This facilitates the storage of a single bit of data (Fig 1 (a)) [7,13]. Following the magnetization switch, the heated region is rapidly cooled under an applied field to store the magnetic polarity of the region. However, in the dual-layer version, the Curie temperature of the bottom layer, $T_{c2}$, is higher than that of the top layer $T_{c1}$ ($T_{c1} < T_{c2}$) [29,30]. In the first write, the laser power is set as high to align the magnetization direction of the bottom layer with the polarity of the write field. Moreover, during the first write, the top layer is inevitably written because of its low $T_c$, resulting in the same magnetization pattern in both layers containing only the up ↑↑ ("1") and down ↓↓ ("-1") magnetization states similar to the single-layer version. Thereafter, in the second write, the laser power is lowered to facilitate writing in only the upper layer without disturbing the magnetization configuration of the bottom layer. This process produces four magnetic states corresponding to ↑↑ ("1"), ↓↓ ("-1"), ↑↓ ("0"), and ↓↑ ("0"), as shown in Fig. 1 (b). These additional magnetization states extend the recording from the conventional 2-levels to 3-levels, and they can potentially be extended to 4-levels if the read head can distinguish the ↑↓ or ↓↑ states.

Experimentally, $T_{c1}$ can be differentiated from $T_{c2}$ either by varying the Fe content in FePt or by using various additives in FePt, such as Cu, Ni, Cr, Mn, B, and Ag [11,12,15,17,29,31-34]. These layers can be magnetically decoupled by introducing a breaking layer between them. In this study, we developed the first experimental synthesis of a multi-level recording medium. Its nanostructure, magnetization-switching mechanism, and read/write performance were investigated using experimental and numerical tools.

**a  Conventional (2-level) recording**

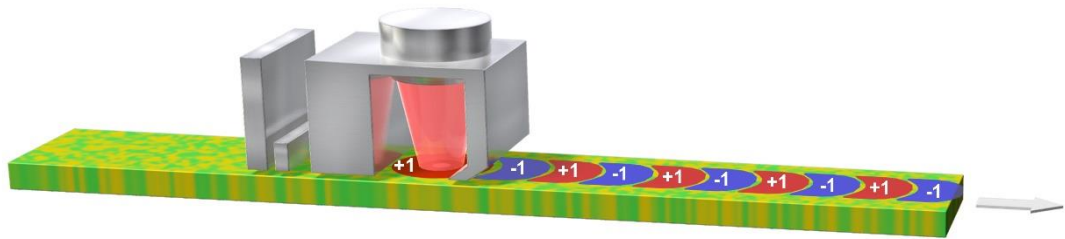

**b  Next-generation (multi-level) recording**

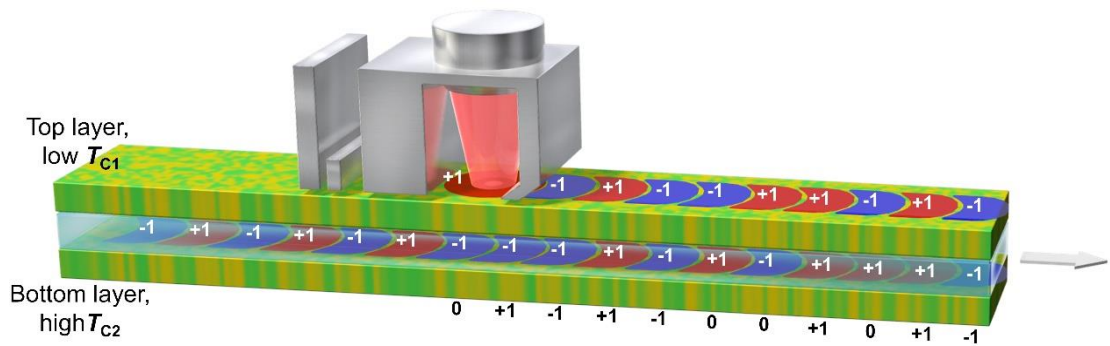

Fig. 1 **Schematic of the concept of multi-level magnetic recording compared to current 2-level magnetic recording technology in hard disk drives.** In both cases, a small area on the recording medium is heated to approximately its $T_c$ to switch the magnetization. The arrow indicates the writing direction. An optical transducer provides a Gaussian heat spot (red color) utilized to heat the area. The magnetic state is stored by cooling the heated area to ambient temperature. **a** In the conventional 2-level recording, the laser power is set as a constant, and magnetization switches by changing the polarity of the write field to obtain a magnetization pattern with (↑ ("1") and ↓("-1")). **b** For multi-level recording, the laser power is tuned along with the polarity of the write field to manipulate the magnetization switching of each layer. During the first write operation, a high laser power is set to heat a small area above $T_{c2}$ of the bottom layer ($T>T_{c2}$) and thus write both layers. During the second write, the laser power is lowered below $T_{c2}$ ($T>T_{c1}$) to write only the top layer. This delivers four different magnetic states (↑↑ ("1"), ↓↓ ("-1"), ↑↓ ("0"), and ↓↑ ("0")) which correspond to a 4-level recording. The recording medium is only a few nm thick, while the breaking layer is half that thickness.

**Experimental**

**A. Breaking layer**

We experimentally investigated various nonmagnetic metallic breaking layers (BL) to magnetically decouple the top and bottom recording layers to realize multi-level recording. An ideal BL should facilitate the growth of the top FePt layer with a smooth and flat surface, high $K_u$, and (001) epitaxy. To select the most appropriate BL, we first modeled the idealized interface between the FePt layers using a continuous film, not a granular one. Consequently, the effects of mismatch strain and surface roughness were minimized, and only the effect of the BL on the magnetic properties, particularly on the switching separation, was investigated. Ru, Cu, Pt, and W were considered for use as a BL owing to their small lattice mismatch (0.5–1.7%) with FePt (Supplementary Fig. S1). High-magnification, cross-sectional high-angle annular dark field (HAADF) and scanning transmission electron microscopy (STEM) images showed that, among these elements, only Ru and Pt exhibited a flat interface with (001) epitaxial growth (Fig. 2 (a,b) and Supplementary Fig. S2). Currently, Pt is not considered because both layers exhibit the same composition, resulting in similar $T_{c1}$ and $T_{c2}$ values in their $M$-$T$ curves. In contrast, compositional differentiation occurs in the case of face-centered cubic (fcc) Ru [35]. The energy-dispersive X-ray spectroscopy (EDS) line scan profile shows the difference in the composition of the top ($Fe_{39}Pt_{61}$) and bottom ($Fe_{42}Pt_{58}$) layers and the diffusion of trace amounts into the bottom within a depth of approximately 1 nm from the Ru layer (Fig. 2(a,c)). This is expected to induce a difference in the $T_c$ values, which is essential for multi-level recording [36]. In addition, despite the predominant (001) epitaxy, the top FePt layer also displayed in-plane variants and a lack of $L1_0$ order during the initial stage of growth (~1 nm) (Fig. 2(a,b)). This affects the effective anisotropy, thus varying the switching field and resulting in a step-like behavior in the Hall bar resistance measurement corresponding to the two coercivity values (Fig. 2(d)). The W and Cu breaking layers formed unexpected (111) epitaxies and island-like microstructures, respectively, as shown in Supplementary Fig. S3.

Although Ru magnetically decoupled the top and bottom layers, it diffused locally into the bottom layer and created local variants and disordered regions in the top layer. Therefore, to understand the comprehensive effect of Ru on the magnetic properties and chemical order, a continuous FePt layer was grown on MgO substrates with Ru contents of $x$ = 0, 0.5, 2, 4, and 8 at.%. Ru substitution decreased the L1$_0$ order while increasing the lattice parameter $c$ (Supplementary Fig S4). Therefore, one reason for the partially disordered region could be the diffusion of Ru into the FePt layers, which results in a different strain state owing to the expansion of the lattice parameter.

The effective anisotropy constant of $K_u$ for $x$ = 0 in Fe$_{42}$Pt$_{58}$ was 3.06 MJ/m$^3$ at 300 K, which is consistent with the literature [37]. Increasing the Ru content to $x$ = 8 in Fe$_{42}$(Pt$_{58-x}$Ru$_x$) decreased $K_u$ to 0.8 MJ/m$^3$ (Supplementary Fig. S5) [34]. Therefore, $K_u$ as well as $H_c$, can be modified in each layer via Ru substitution or diffusion. Further, $T_c$ decreased from 610 K for Fe$_{42}$Pt$_{58}$ to 540 K via the substitution of a small amount of Ru in Fe$_{42}$Pt$_{57.5}$Ru$_{0.5}$ (Supplementary Fig. S6 (a,b)) [34,36,38]. The modification of $K_u$ and $T_c$ via Ru introduction benefits multi-level recording by controlling the magnetization reversal in each layer if the Ru substitution remains at $x \leq 0.5$ at.%.

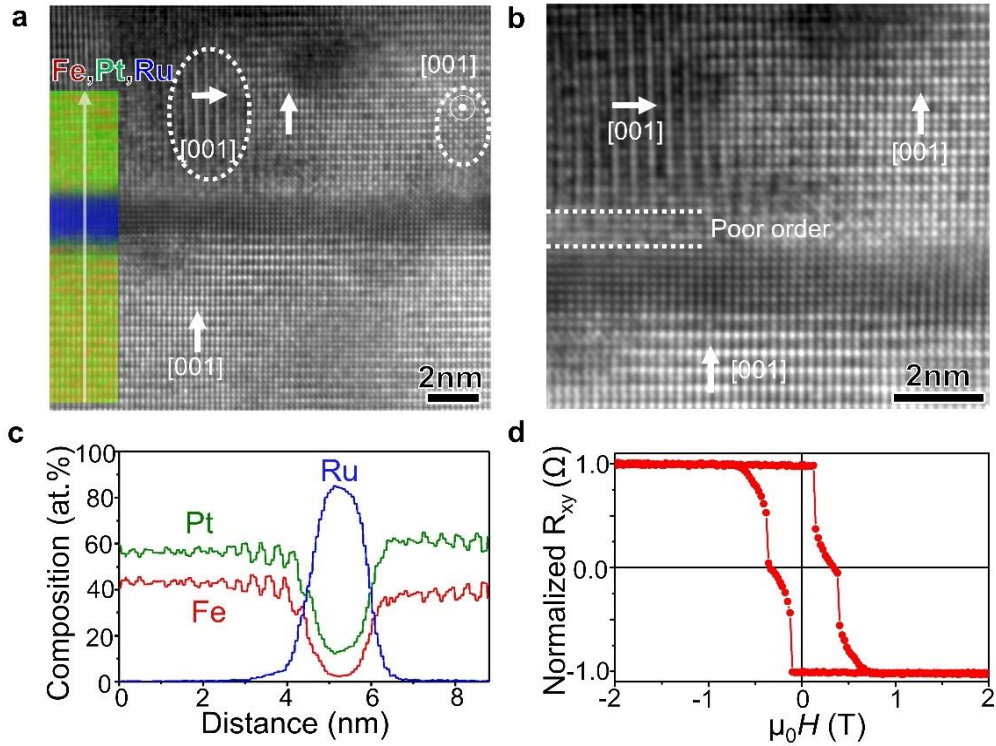

Fig. 2. **Microstructural features of FePt/Ru/FePt continuous film. a, b** HAADF and STEM-EDS maps of Fe, Pt, and Ru with **c** line scan for FePt (17 nm)/Ru (2 nm)/FePt (17 nm). **c** EDS line scan reveals the difference in the composition of the top and bottom layers, with a trace amount of Ru diffusion into the bottom. **d** Transverse normalized Hall-bar resistance ($R_{xy}$) measurement as a function of the external magnetic field, where $R_{xy}$ corresponds to the out-of-plane of the magnetization, showing that each layer has a different switching field.

### B. Dual nanogranular magnetic layers

Dual nanogranular magnetic layers as a multi-level recording medium was obtained by growing a (001)-textured FePt-20 vol.% C granular bottom layer (4.5 nm) on a MgO (001) substrate, followed by a Ru-C breaking layer (2.9 nm) and FePt-C granular top layer. This was confirmed via the cross-sectional HAADF-STEM image and EDS map in Fig. 3 (a,b). Carbon is preferred as a segregant to reduce the average grain size and maximize the grain density while maintaining the $L1_0$ order [14,39,40]. The FePt grains primarily exhibited a columnar shape with a relatively flat surface and certain curved grains. In Fig. 3(c), the high-magnification HADDF image reveals that the FePt grains in the bottom layer exhibit mainly well L1$_0$-ordered (001) texture with some in-plane variants. In contrast, the FePt grains in the top layer exhibit more in-plane variants and partially disordered regions. The EDS maps and line scans in Fig. 3(e-g) show that Ru diffused to both layers and to the intergranular region. To investigate the influence of the Ru-C BL on the degree of the L1$_0$ order, the XRD patterns of the single- and dual-layer media were evaluated. The large chemical order parameter, S = 0.8, of the single FePt-C layer decreased to 0.63 with the growth of the BL and top layer (Supplementary Fig. S7). Another factor that disturbed the (001) texture was the shape of FePt grains. Interfacial strain is required to suppress the formation of in-plane variants, as is the case for the FePt/MgO interface in the bottom layer [41,42]. However, such strain may lack at the curved Ru-FePt interface, leading to an increase in the number of in-plane variants in the top layer.

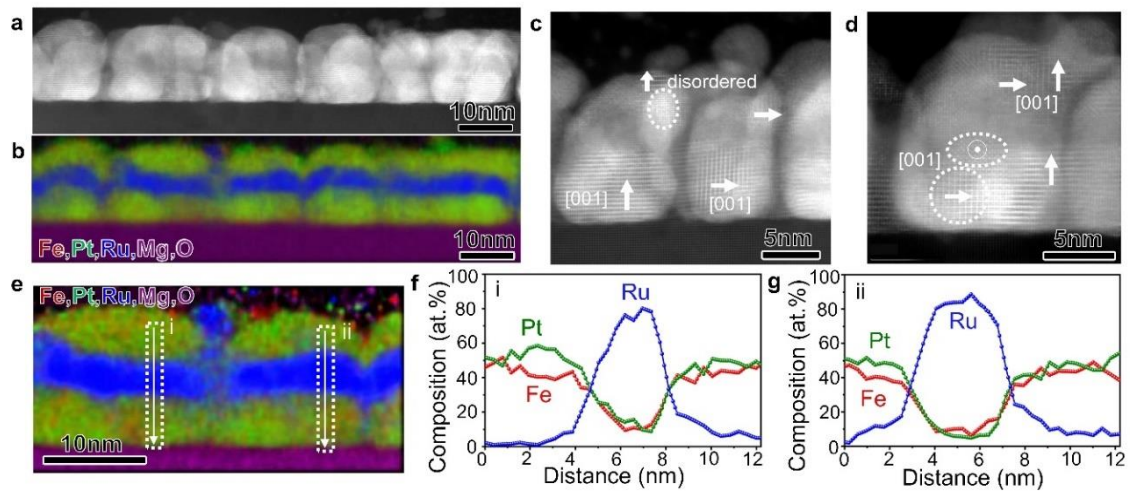

Fig. 3 **Overall nanostructure of the developed media for multi-level magnetic recording. a** HADDF image **b** EDS map for FePt-20 vol.% C/Ru- 20 vol% C/FePt-20 vol% C of granular dual layer media shows that FePt-C layers are separated by using a Ru-C breaking layer. **c, d** High magnification HADDF images indicate that the additional in-plane variants are present on the top layer, in addition to the (001) texture. **e** High magnification EDS map with **f, g** its line scan revealing that Ru diffuses to both the layers and to the intergranular region.

The average grain size of the media was obtained as 12.7± 5.7 nm from the top-view bright-field transmission electron microscopy (TEM) images in Fig. 4(a). Fig. 4(b, c) show the experimental out-of-plane (OOP) and in-plane (IP) magnetization curves, along with the simulated curves. The IP curve indicates strong perpendicular magnetic anisotropy. Based on half of the mid-distance between the saturation and plateau regions of the OOP curve, the coercivities of the top ($H_{c1}$) and bottom ($H_{c2}$) layers were determined to be $\mu_0 H_{c1} = 0.24$ T and $\mu_0 H_{c2} = 3.82$ T. The large $\mu_0 H_{c2}$ value confirms good epitaxial growth of the highly ordered FePt (001). To interpret the magnetic properties of the obtained two-layer FePt films and distinguish them from the individual layers, the measured OOP hysteresis loops were analyzed using micromagnetic simulations. We reproduced the microstructural features of the films using a finite element model to perform micromagnetic approximations of the demagnetization curves, where the mean magnetic anisotropy constants ($K_{top}$ and $K_{bot}$), their standard deviations, and the volume fractions of the in-plane variants were used as approximation parameters (Supplementary Fig. S8). The simulated $M$-$H$ curve for $K_{bot} = 2.70 \pm 0.45$ MJ/m$^3$ ($K_{top} = 0.6 \pm 0.1$ MJ/m$^3$) and for 8 vol.% (30 vol.%) of in-plane variants in the bottom (top) layer is consistent with the experimental ones. An increase in the number of in-plane variants on the top layer was observed using high-resolution cross-sectional TEM, resulting in a decrease in $K_{top}$. This was attributed to the distribution of the L1$_0$ order and Ru diffusion (Fig. 3 (c)). Based on the micromagnetic simulation, the in-plane variants of the top layer had minimal impact on the shape of the demagnetization curve. However, increasing $K_u$ while decreasing the variants enhanced remanence ($\mu_0 M_r$) and eliminated the plateau region (Fig. 4 (c)). Nevertheless, the high slope d$M$/d$H$ clearly indicates a two-layer recording system with distinct $K_u$ of top and bottom layers.

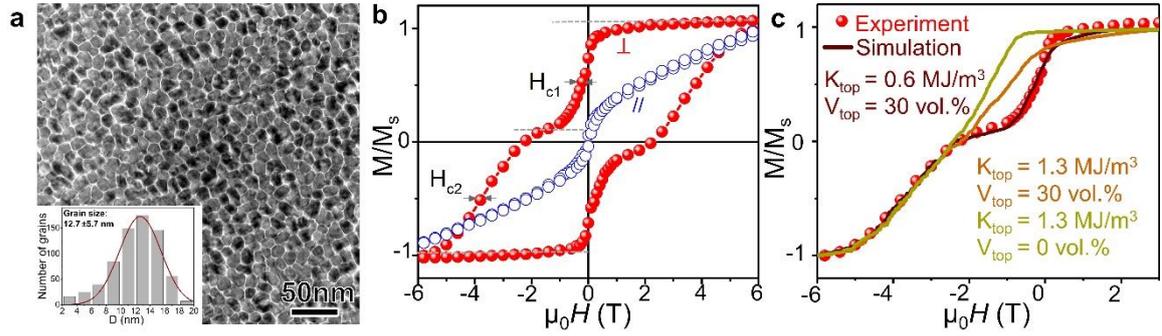

Fig. 4. **Microstructure of granular FePt-C/Ru-C/FePt-C films and corresponding magnetization curves. a** In-plane bright-field TEM images, with an inset showing the grain size distribution, reveal that the average grain size is 12.7± 5.7 nm. **b** Corresponding out-of-plane (⊥) and in-plane (//) magnetization curves show that $\mu_0H_{c1}$= 0.24 T and $\mu_0H_{c2}$ = 3.82 T. **c** The simulated (line) and experimental (circle) *M-H* curves are shown together, along with the variation of the simulated *M-H* curve with an anisotropy constant $K_{top}$ and a variant $V_{top}$.

**Temperature-dependent magnetic measurement and writing simulation**

The $T_c$ of each layer in the dual nanogranular magnetic layers was estimated from *M-T* measurement at 0.1 T in Fig. 5(a), and it was found to be $T_{c1}$ = 526 K and $T_{c2}$ = 620 K. $T_c$ can be affected by various factors such as Ru diffusion, deviation in the Fe and Pt contents of the FePt grains, and the L1$_0$ order parameter. Therefore, $T_{c1}$ = 526 K may belong to the top layer, which contains most of these factors.

The writing performance of the HAMR media was investigated using temperature-dependent magnetic measurements (Fig. 5(b)). Unlike in the HAMR writing procedure, in this magnetic writing test, the medium was heated as a whole, rather than locally. First, the medium was saturated, corresponding to the magnetic state of ↑↑ ("1"), under 7 T at 300 K. To switch the top layer with a low $K_{top}$ (0.6 MJ/m$^3$) and $T_{c1}$ (526 K), a small reverse field ($\mu_0H$ = -0.1 T) was applied while increasing the temperature from 300 K to $T_{an}$ = 400, 500, 600, and 700 K. The medium was then cooled to 300 K to store the obtained magnetic state. Consequently, the magnetization of the top layer was reversed, and *M* = 0 was obtained by cooling the medium from $T_{an}$ = 600 to 300 K, a temperature higher than $T_{c1} = 526$ K, $T_{an} > T_{c1}$. This process delivered a magnetic state of ↓↑ ("0"). For switching the bottom layer with high $K_{bot}$ (2.7 ±0.45 MJ/m³), the reverse field was increased to $\mu_0 H = -1$ T, and the same process was followed. The magnetization of the bottom layer was reversed by cooling from 700 to 300 K, a temperature higher than $T_{c2} = 610$ K, $T_{an} > T_{c2}$. This delivered the magnetic state of ↓↓ ("-1").

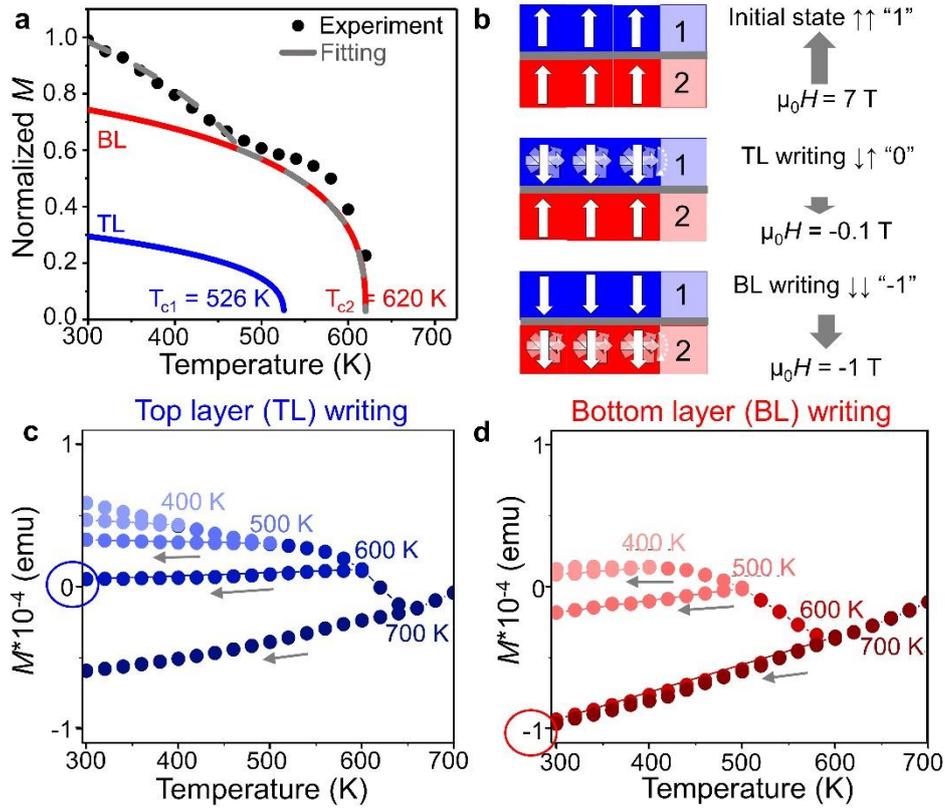

Fig. 5. **Temperature-dependent magnetic measurement of granular FePt-C/Ru-C/FePt-C media. a** *M-T* measurement is performed under 0.1 T wherein the top and bottom layers exhibit $T_{c1}$ of 526 K and $T_{c2}$ of 620 K. **b** Schematic representation of magnetic measurement to mimic the writing procedure. **c** *M-T* measurement under an applied field of $\mu_0 H = -0.1$ T to switch the magnetization of the top layer at 600 K corresponding to the ↓↑ state ("0"). **d** Increasing $\mu_0 H = -1$ T to switch the magnetization of the bottom layer corresponding to ↓↓ ("-1").

In addition, a multi-level HAMR recording simulation was performed based on the experimentally measured parameters. A heat spot with a Gaussian intensity distribution and globally uniform write field (0.2 T was used to heat the media and switch the magnetization in the desired direction as the media cooled down). Fig. 6 (a, b) shows images of the top and bottom layers at the end of a two-pass writing, which is described in "Concept of multi-level recording in HAMR." Writing started with an AC-erased medium corresponding to each grain in a magnetically randomly oriented state. The first pass was written at 680 K, which is higher than the $T_c$ of both layers, and a bit length of 50 nm. This resulted in the successful writing of both layers but differing track widths, which were narrower (≈50 nm) and wider (≈100 nm) for the bottom and top layers, respectively, owing to the different $T_c$ ($T_{c1}$<$T_{c2}$). For the second pass, the laser power was lowered to 540 K, below $T_{c2}$, and the bit length was increased to 100 nm to write only the top layer. However, remnants of the first track remained at the track edge of the top layer, and the magnetization within the written bits was not fully saturated. This is attributed to certain grains switching in the direction opposite to that of the write field during cooling, which is used to store the magnetic state. This problem can be solved by either increasing the write field or increasing $K_u$ of the top layer. Fig. 6 (c) shows the magnetization along the down-track direction in the top and bottom layers, averaged over a 40 nm track width, along with the sum of the magnetization of the two layers. Herein, four different magnetization states were observed. The magnetization corresponding to ↑↓ and ↓↑ was M≠0, which was easily distinguished. Further, the recording can be expanded from 3-levels to 4-levels.

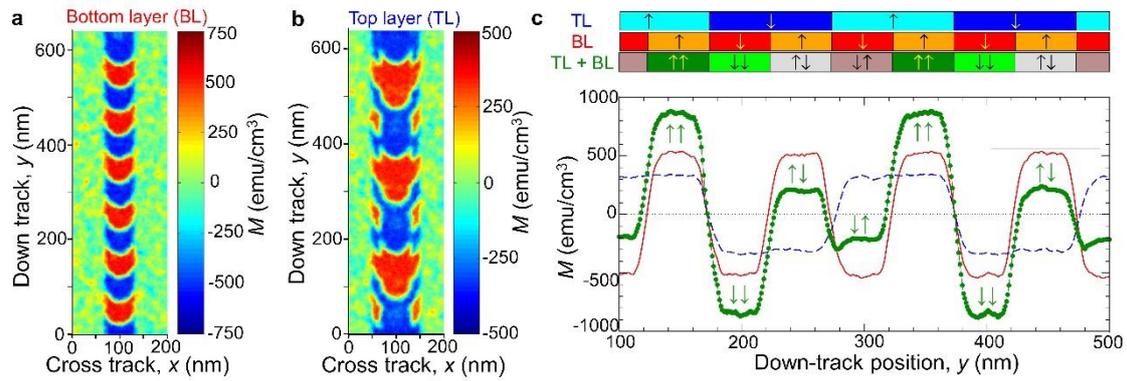

Fig. 6. **Multi-level HAMR recording simulation for granular FePt-C/Ru-C/FePt-C media** Average magnetization patterns of 50 tracks after a two-pass write. **a** Pass 1: Writing on the bottom layer with $T_{max}$ = 680 K, bit length = 50 nm. **b** Pass 2: Writing on the top layer with $T_{max}$ = 540 K **c** Average magnetization along the down-track direction for the top and bottom layers. The magnetization of the top and bottom layers and the sum of the magnetization of both layers are indicated by a blue dash line, red solid line, and green circle, respectively.

## Conclusion

Proof of concept in the multi-level recording has been developed for the first time successfully by separating FePt-C (20 vol.%) nanogranular layers with a cubic Ru-C breaking layer. Both layers exhibited mainly a $L1_0$ ordered [001] texture along with certain in-plane variants based on micromagnetic simulation with a finite element model reproduced from experimental microstructural features. The higher variants in the top layer are attributed to the Ru diffusion and the spherical shape of FePt grains, resulting in a low $\mu_0 H_{c1}$ and $K_{top}$ compared to the bottom layer. The $T_c$ of each layer differed owing to Ru diffusion, deviations in the Fe and Pt content of the composition, and the order parameters, which facilitates multi-level recording.

The writing performance of the media was evaluated using experimental magnetic measurements and a HAMR recording simulation. Both layers were written successfully following the first pass; however, during the second pass, certain remnants of the first track remained in the top layer. This can be overcome by increasing $K_{top}$ by improving the degree of the $L1_0$ order and by searching for a suitable spacer layer. After the two-pass writing, the magnetic states of ↑↑, ↓↓, and ↓↑ were obtained, with ↑↓ and ↓↑ having M≠0, thus, allowing them to be distinguished and enabling an expansion from 3-level to 4-level recording. For this concept, due to the magnetic field strength of the magnetic head and the Curie temperature requirements, the number of layers is estimated to be three at most. Further optimization of microstructure and magnetic properties would pave the way to 10 Tbit/in$^2$.

**Methods**

**Experimental methods**

All the films were prepared using an ultrahigh-vacuum co-sputtering system with a base pressure of approximately 10$^{-7}$ Pa. To investigate the breaking layer, FePt (17 nm)/X (with a gradual thickness (2–7 nm)/FePt (17 nm)) was deposited on an STO (001) substrate at 500 ºC for X = Ru, Cu, and Pt. The dual-layer films were patterned into Hall bars for transverse Hall bar resistance measurements. An alternating MgO/FePt-C film stack (20 vol.%) (4.5 nm)/FePt(0.26 nm)/Ru-C (20 vol.%) (2.9 nm)/Ru (0.11 nm)/FePt-C (20 vol.%) (4.5 nm)/C-cap was grown at 500 ºC as a granular dual medium. The average heights of the layers (FePt-C (5.5±0.7 nm)/Ru-C (3.0±0.6 nm)/ FePt-C (5.1±0.6 nm)/) obtained from the EDS maps were consistent with the design. Thin FePt and Ru layers were deposited to prevent C migration. A C capping layer was deposited at 300 K to prevent surface oxidation. MgO single-crystal substrates are preferred for promoting a strong [001] texture and minimizing misorientation in (001)-textured FePt grains [9,42]. The film stack of MgO/FePt-Ru (30 nm)/C-cap (4.5 nm) was grown at a substrate temperature of 500 ºC to perform the comparisons. X-ray diffraction (XRD) Rigaku

SmartLab with a Cu X-ray source was used for crystallographic and degree of $L1_0$ order analyses. Electron-transparent thin specimens were prepared for transmission electron microscopy (TEM) using a focused ion beam lift-out technique (FEI Helios Nanolab 650). To prevent damage during the ion beam milling, the film surfaces were coated with Ni. TEM was performed using a Titan G2 80–200 with a probe aberration corrector. Energy-dispersive X-ray spectroscopy (EDS) was conducted using an FEI Super-X EDX detector. A 7 T Quantum Design superconducting quantum interference device magnetometer (SQUID) was used for the magnetic measurements. The films were embedded in high-temperature cement for the measurements at temperatures above 380 K. The Curie temperature ($T_c$) was determined by the least squares fitting using Kuz'min's equation [43]. The magnetization direction ($\theta_M$) dependences of torques were measured using the anomalous Hall effect to deduce the first- and second-order uniaxial anisotropy constants $K_1$ and $K_2$ as described by Ono et al. [41].

**Micromagnetic Approximation of Hysteresis Loops:**

A finite element model of a two-layer FePt granular medium was developed using Voronoi-based tessellation. The bottom layer was represented by 5 nm thick columnar grains with a top-view grain size of 12.5 (2.0) nm, which reproduced the experimental grain size distribution. The top layer with a thickness of $t_{top}$ was fabricated by the out-of-plane translation of the bottom layer through a 3-nm-thick nonmagnetic gap (breaking layer). Following the TEM observations of the real samples, in-plane variants were introduced into the model by randomly splitting certain grains into several regions with mutually orthogonal easy magnetization axes. The total volume fractions of the in-plane variants were controlled in both layers using $V_{top}$ and $V_{bot}$. The magnetic anisotropy constant was distributed among the grains, with mean values of $K_{top}$ and $K_{bot}$ as well as standard deviations that were individually prescribed for the top and bottom layers, respectively. The saturation magnetization and exchange stiffness values were assumed to be the same for all grains, that is, $\mu_0 M_s$ = 1.43 T and $A$ = 10 pJ/m, respectively [39]. Micromagnetic simulations were performed using the FastMag software by solving the Landau–Lifshitz–Gilbert equation with a damping constant α = 1 [42]. The parameters $t_{top}$, $V_{top}$, $K_{top}$, $V_{bot}$, and $K_{bot}$ were varied in a grid-search manner to approximate the experimental hysteresis loops.

**Multilevel HAMR recording simulation:**

The model uses the Landau–Lifshitz–Bloch equation, which is suitable for modeling the behavior of magnetic materials at temperatures up to and beyond $T_c$ [43]. The temperature dependence of $M_s$ was calculated by assuming a Brillouin function with $J$ = 1, and $K_u$ was derived from $M_s$ as $K_u(T)$ μ $M_s(T)^2$ [44]. Each image depicts the average magnetization of 50 tracks, where a magnetic head was used to read or write data. The simulation parameters used were the experimental $T_c$ and predicted $K_u$ based on the experimental $M$-$H$. The recording medium had an average grain size, average grain pitch, and grain size distribution of 13 nm, 14 nm, and 9%, respectively. The top and bottom layers were both 6 nm thick, separated by a 3 nm nonmagnetic layer. The heat spot was moved along the medium at a velocity of 4 m/s. The ambient temperature was set at 300 K. The heat spot had a two-dimensional Gaussian temperature distribution in the media with a half width at half maximum (HWHM) of 50 nm, that is, 50 nm from the center of the heat spot, and the temperature was 300 + (0.5×($T_{max}$ - 300)) K where $T_{max}$ is maximum temperature. The tracks showed a signal-to-noise ratio closest to the average of the 50 written tracks. Thus, reversed grains within written bits are more likely to occur in the top layer.

**Acknowledgments:** This work was supported in part by the JST-CREST (JPMJC22C3) and MEXT program: Data Creation and Utilization-Type Material Research and Development Project (JPMXP1122715503).

# Supplementary Information

P. Tozman[1*], S. Isogami[1], I. Suzuki[1], A. Bolyachkin[1], H. Sepehri-Amin[1], S.J. Greaves[2]
H. Suto[1], Y. Sasaki[1], H.T. Y. Chang[3], Y Kubota[3], P. Steiner[3], P.-W. Huang[3], K. Hono[1]
and Y.K. Takahashi[1*]

[1] National Institute for Materials Science, Tsukuba, Ibaraki 305-0047, Japan
[2] Research Institute of Electrical Communications Tohoku University, Sendai, Japan
[3] Seagate Technology, Recording Media Organization, Fremont, CA. 94538 USA

## A. Magnetic properties influenced by various breaking layers

Fig. S1(a) shows the film stack of FePt (17 nm)/breaking layer (*X*) (2–7 nm)/FePt (17 nm), which was used to investigate the *X*-dependent magnetic properties. This stack was deposited on an STO (001) substrate at 500 ºC for *X* = Ru, Cu, Pt, and W. The lattice parameters of these elements are presented in Fig. S1(b) [1,2]. Fig. S1(c) shows the transverse resistance ($R_{xy}$) of the Hall bar devices as a function of the magnetic field applied along the film normal ($\mu_0 H$), where $R_{xy}$ was normalized by the value for $\mu_0 H = 2$ T.

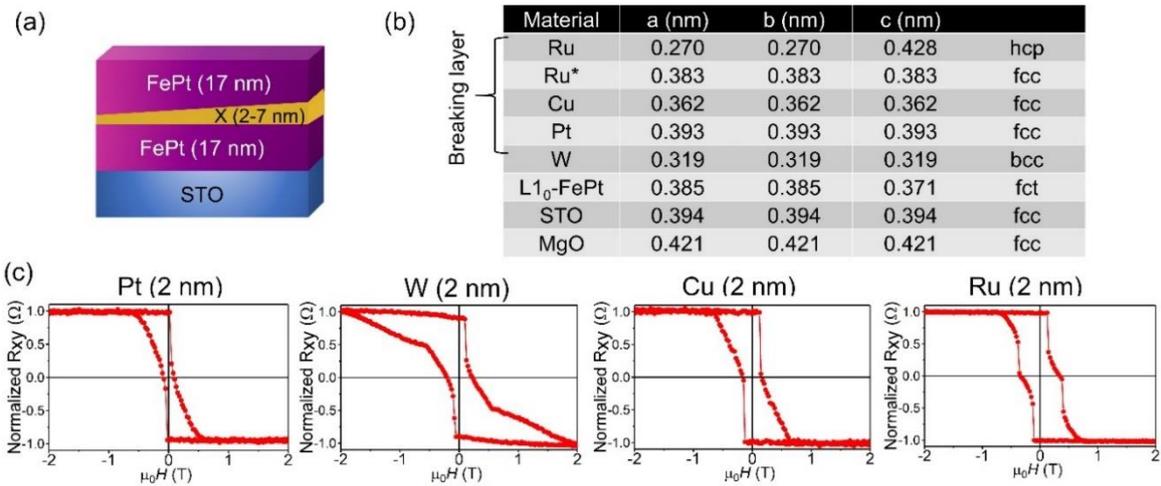

Fig. S1. (a) Stacking structure with highly textured FePt films grown on an STO (001) substrate. (b) Lattice parameters of various breaking layers (*X*), FePt, and substrates. Note that the * symbol indicates Ru with the fcc structure (c) Normalized transverse resistance ($R_{xy}$) of the Hall bar devices with the 2 nm-thick *X* = Pt, W, Cu, and Ru.

## B. Structural analysis using transmission electron microscopy

Fig. S2 shows the nanobeam electron diffraction patterns along with high-magnification, cross-sectional high-angle annular dark-field (HAADF) scanning transmission electron microscopy (STEM) images and energy-dispersive X-ray spectroscopy (EDS) line for FePt (17 nm)/Pt (2 nm)/FePt (17 nm). The bottom and top FePt layers exhibit (001) epitaxy, that is, fine crystallographic orientation along the [001] direction (Fig. S2(a,b)). Both layers were composed of $Fe_{40}Pt_{60}$, which was slightly richer in Pt than the single FePt layer with $Fe_{42}Pt_{58}$ (Fig. S2(c)). Because both layers exhibit the same composition, the $M$-$T$ curve, measured under $\mu_0 H = 0.1$ T, shows a single $T_c$ resulting in similar $T_{c1}$ and $T_{c2}$ values (Fig. S2(e)). Therefore, Pt can be considered a spacer for granular media if $T_{c1}$ and $T_{c2}$ are differentiated in the future.

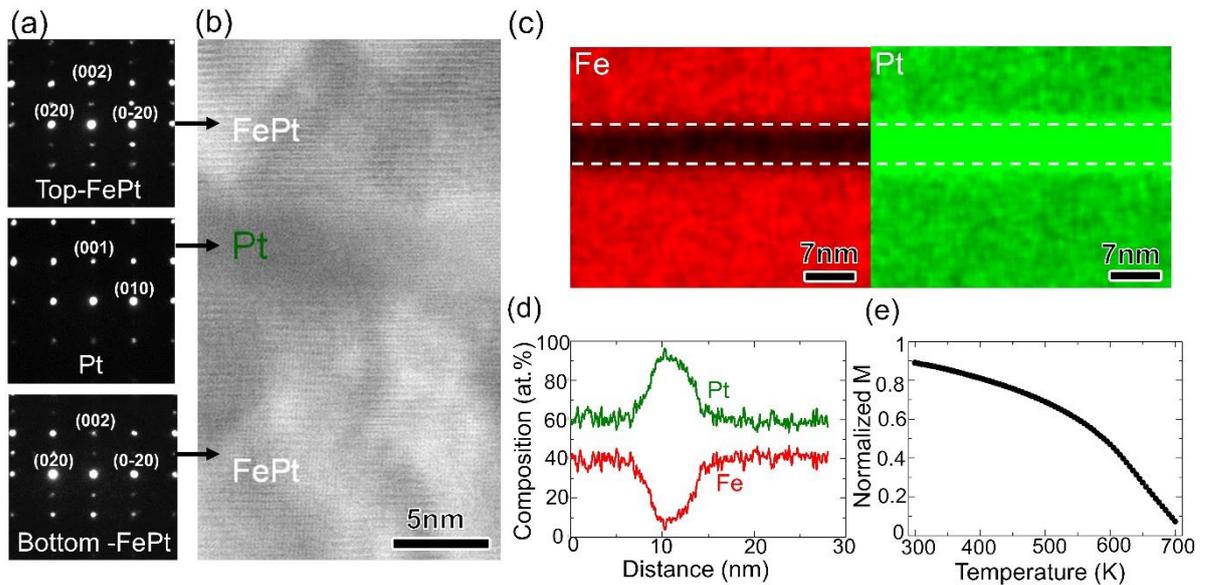

Fig S2. (a) Nanobeam electron diffraction patterns with FePt-[100] zone axis for the bottom- and top-FePt layers. (b) High-resolution HAADF-STEM images. (c) STEM-EDS maps of Fe and Pt along with (d) line scan. (e) Temperature-dependent normalized magnetization, $M$-$T$ curve, for the FePt/Pt/FePt structure.

In Fig. S3, the nanobeam electron diffraction patterns along with high-magnification cross-sectional HAADF-STEM images and EDS maps are shown for FePt (17 nm)/W (2 nm)/FePt (17 nm). The bottom FePt and W-breaking layers exhibit (001) epitaxy. Although the bottom interface between the bottom FePt and W is smooth, that between the top FePt and W is rough, which causes (111) growth in the top FePt.

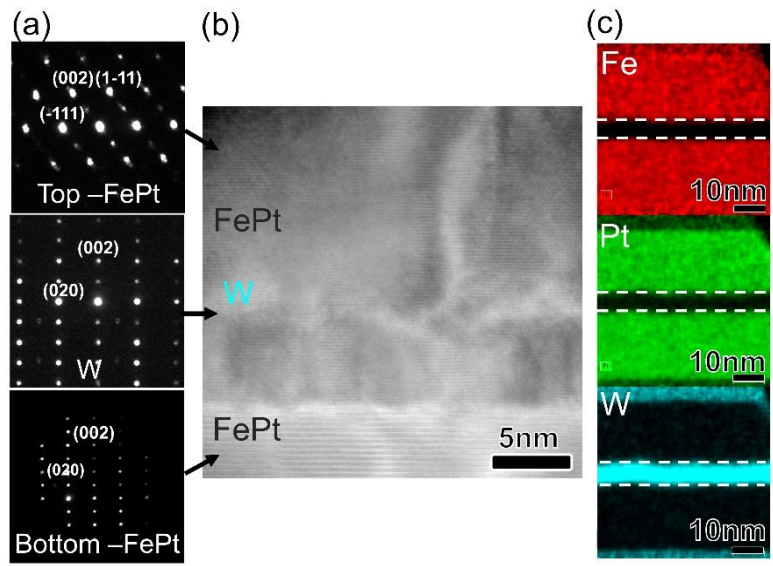

Fig. S3. (a) Nanobeam electron diffraction patterns of bottom-FePt, W breaking layer with [110] zone axis, and top-FePt with zone axis [110]. (b) High-resolution HAADF-STEM images. (c) STEM-EDS maps for the FePt/W/FePt structure.

### C. Pt substitution by Ru

Fig. S4(a–c), show the XRD patterns, atomic order parameter $S$, and lattice parameter $c$, respectively, as functions of the Ru content in $Fe_{42}(Pt_{58-x}Ru_x)$. The peaks observed at approximately 23° and 48° originate from the superlattice 001 and fundamental 002 planes, respectively. The $S$ parameter of FePt was consistent with the value reported in the literature for a given substrate temperature of 500 ºC [3]. Further, the Ru substitution decreased the $L1_0$ order while increasing the lattice parameter $c$.

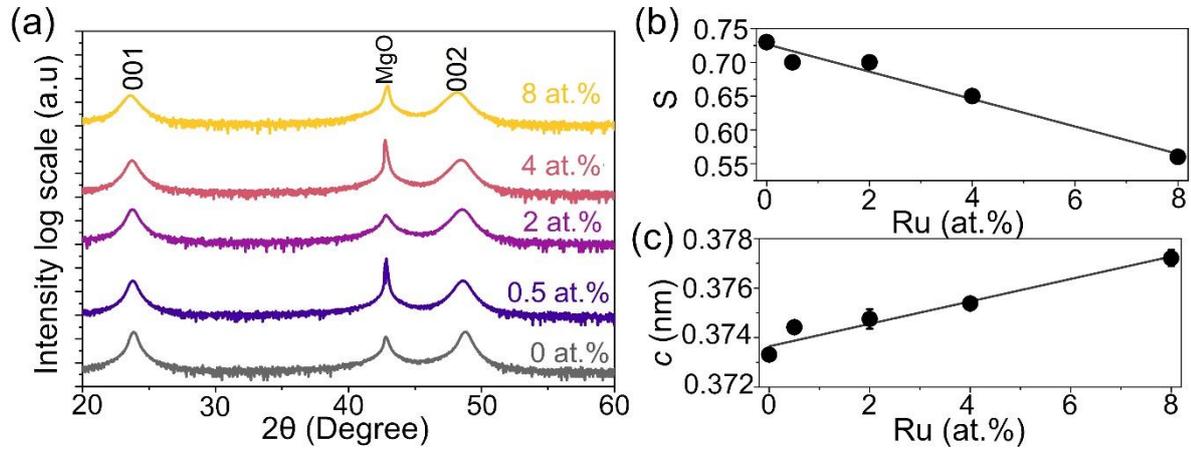

Fig. S4. (a) Out-of-plane X-ray diffraction (XRD) patterns for the $Fe_{42}(Pt_{58-x}Ru_x)$ films with $0 \leq x \leq 8$ along with the Ru content dependence of (b) the atomic order parameter $S$ and (c) the lattice parameter $c$.

The magnetocrystalline anisotropy constants $K_1$ and $K_2$ were deduced using a torque measurement method based on the anomalous Hall effect (AHE) in the temperature range of 100–370 K [4] (Fig. S5 (a)). Although the temperature dependence of $K_1$ follows the same trend as in the literature, $K_2$ has an opposite sign to the reported value [5]. This discrepancy may be because of the different analysis techniques: the Sucksmith Thompson (ST) and AHE torque methods used for determining $K_1$ and $K_2$, wherein the ST method induces a large error, particularly at $K_2$ [6]. Increasing the Ru content in $Fe_{42}(Pt_{58-x}Ru_x)$ decreased $K_1$ and changed the sign of $K_2$ from negative to positive.

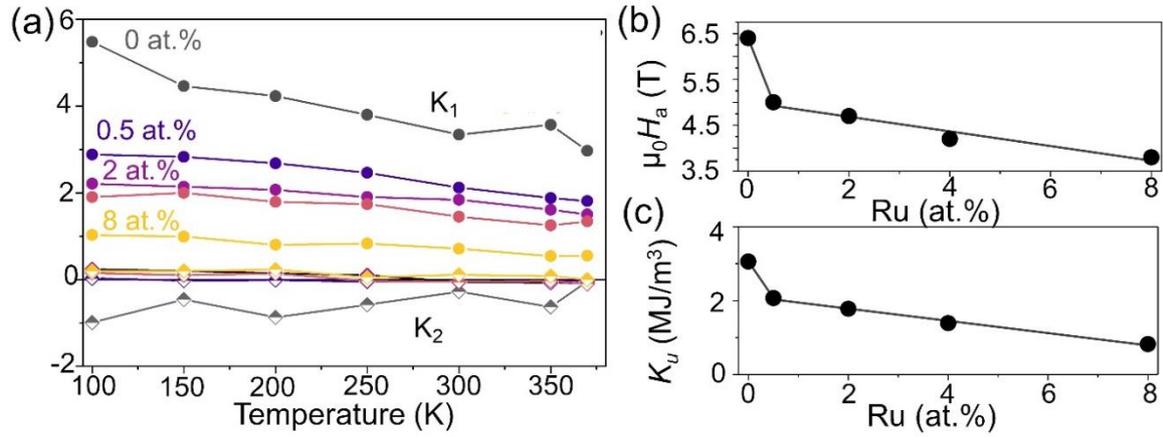

Fig. S5. (a) Temperature dependence of magnetocrystalline anisotropy constants of $K_1$ and $K_2$ in the Fe$_{42}$(Pt$_{58-x}$Ru$_x$) films with $0 \leq x \leq 8$ at.%. Filled and half-filled marks correspond to $K_1$ and $K_2$, respectively. Ru content dependence of (b) the anisotropy field ($\mu_0 H_a$) and (c) the effective anisotropy constant, $K_u = K_1 + K_2$

Fig. S6(a) shows the saturation magnetization versus temperature for Fe$_{42}$(Pt$_{58-x}$Ru$_x$), where the circle and line indicate the experimental and fitted data, respectively. The Curie temperature ($T_c$) was determined by the least squares fitting using Kuz'min's equation [7]. $T_c$ of Fe$_{42}$Pt$_{58}$ decreased from 610 to 540 K upon substituting only a small amount of Ru in Fe$_{42}$Pt$_{57.5}$Ru$_{0.5}$. A further increase in Ru content was expected to further decrease $T_c$, following a linear trend with a smaller decrement rate of ≈15%/at.%, which is close to the rate reported in the literature, i.e., ≈17%/at.% (Fig. S6 (b)) [8]. The decrease in $T_c$ was followed by a decrease in $M_s$, as shown in Fig. S6(c), which has been confirmed via DFT calculation [9].

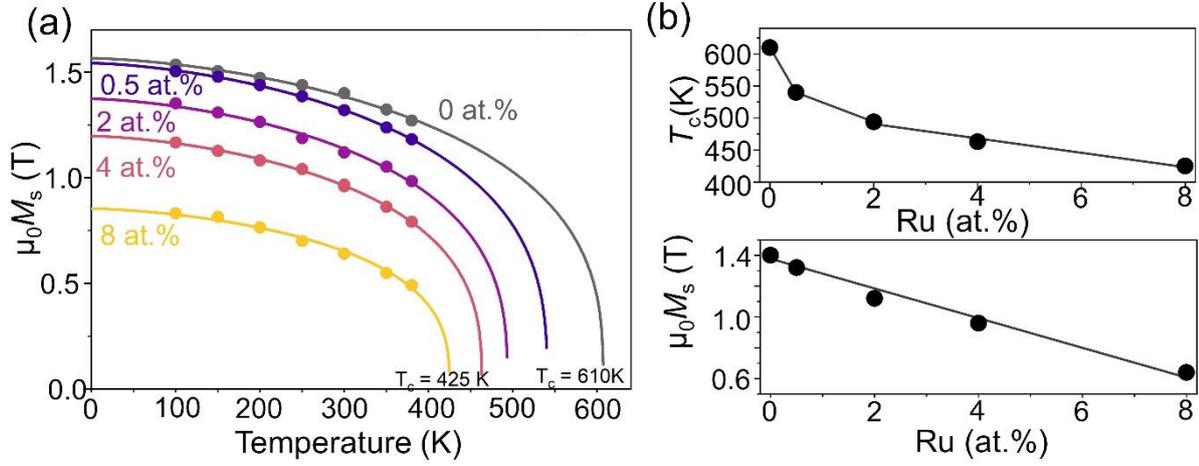

Fig. S6. (a) Temperature-dependent saturation magnetization ($\mu_0 M_s$) for the $Fe_{42}(Pt_{58-x}Ru_x)$ with $0 \leq x \leq 8$ at.%, where the dots and lines indicate experimental and fitted data, respectively. Ru content dependence of (b) the Curie temperature ($T_c$) and (c) the saturation magnetization ($\mu_0 M_s$).

The chemical order of Fe and Pt in the FePt grains of the FePt-C granular layer was estimated using $S = 0.85 \times \sqrt{I_{001}^{exp}/I_{002}^{exp}}$ [10]. A large chemical ordering parameter, $S = 0.8$, was observed, even at a thickness as small as 4.5 nm (Fig. S7(a)). This value is comparable to that reported in the literature for a single FePt-C granular layer with a thickness of 8 nm [11,12]. However, $S$ decreased to approximately 0.63 in the stacking with dual FePt-C layers, as shown in Fig. S7(b), which can be attributed to the lower $S$ in the top layer.

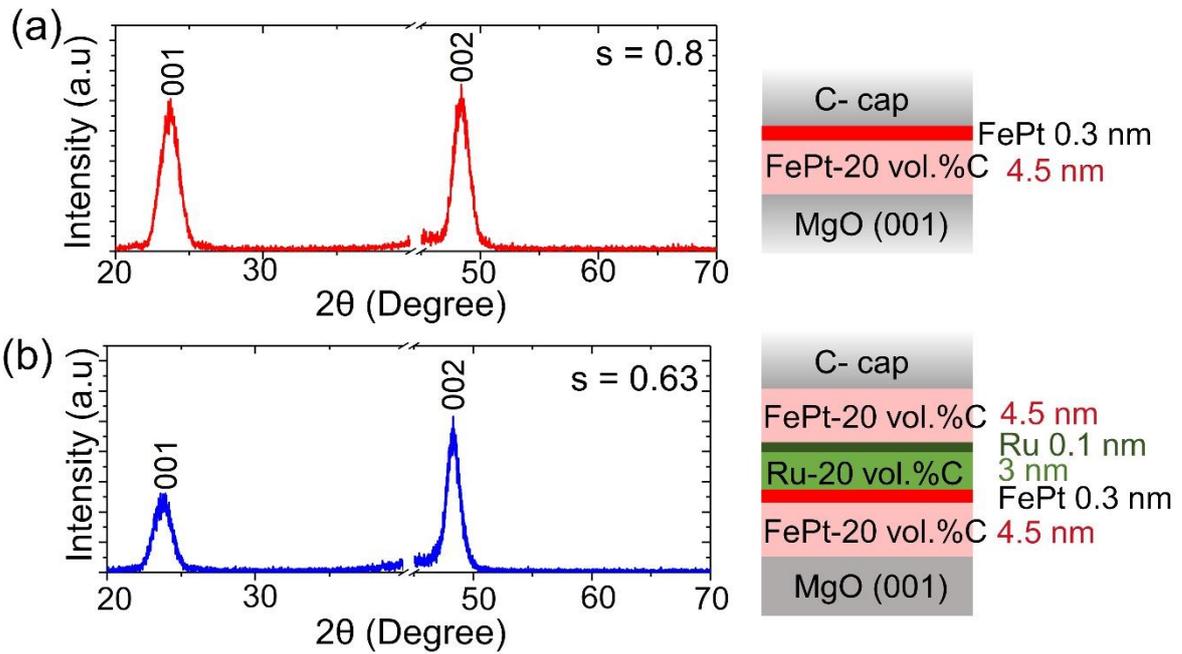

Fig. S7. Out-of-plane XRD patterns and estimated chemical order parameters (*S*) for the stacking with (a) single FePt-C nanogranular layer and (b) dual layers.

### D. Micromagnetic simulation for the dual nanogranular films

In Fig. S8(a), the microstructural features of the dual-nanogranular film were reproduced using a finite element model based on the experimental microstructure. The table in Fig. S8(b) shows the estimated magnetic anisotropy constant ($K_u$), volume fraction of the in-plane variants ($V$), and height of the layer ($t$) for the top and bottom layers from the micromagnetic approximation. As the actual thickness of the top layer may deviate from the nominal thickness during deposition on the curved interface of the bottom layer, this study introduced the thickness of the top layer as an approximation parameter.

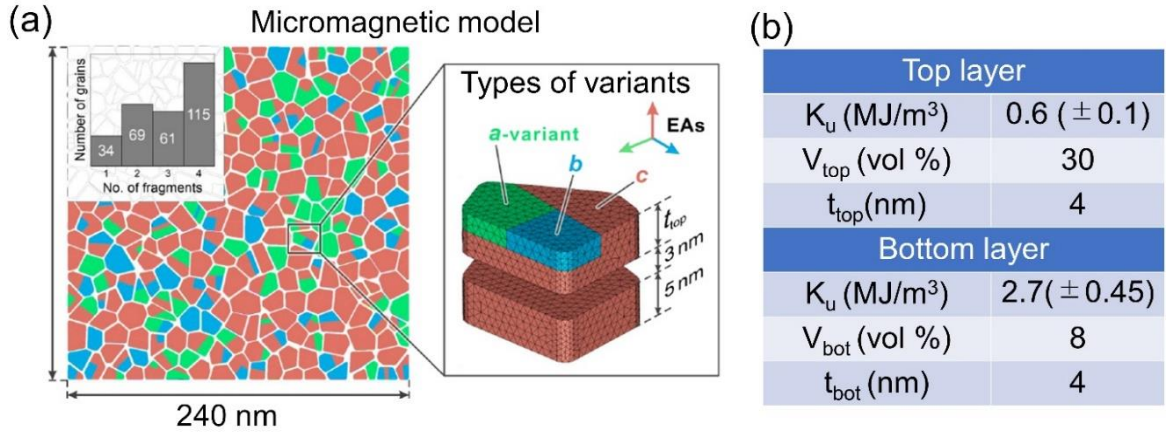

Fig. S8. (a) The micromagnetic model along with (b) estimated parameters for the FePt-20 vol.%C/Ru- 20 vol% C/FePt-20 vol% C.